\pdfoutput=1

\documentclass[conference]{IEEEtran}

\renewcommand\IEEEkeywordsname{Keywords}

\IEEEoverridecommandlockouts
\usepackage{cite}
\usepackage{array}
\usepackage{listings}

\usepackage[misc]{ifsym}
\usepackage{amsmath}
\usepackage{url}
\usepackage[pdftex]{graphicx}
\usepackage{epstopdf}
\usepackage{hyperref}
\usepackage{cleveref}
\usepackage[english]{babel}

\newcommand{\XeonPhi}{Xeon Phi }
\newcommand{\Xeon}{Xeon } 
\newcommand{\Intel}{Intel } 
\newcommand{\VTune}{VTune } 

\usepackage{etoolbox}

\makeatletter
\apptocmd\@makecaption{\par}{}{%
  \errmessage{\noexpand\@makecaption could not be patched}%
}
\makeatother

\begin{document}


\title{Performance Optimisation of Smoothed Particle Hydrodynamics Algorithms for Multi/Many-Core Architectures}

\author{\IEEEauthorblockN{Fabio Baruffa,\texorpdfstring{\textsuperscript{(\Letter)}},
Luigi Iapichino,\texorpdfstring{\textsuperscript{(\Letter)}},  
Nicolay~J.~Hammer and 
Vasileios Karakasis$^*$\thanks{$^*$Current address: CSCS - Swiss National Supercomputing Centre, Lugano, Switzerland}} 
\IEEEauthorblockA{Leibniz Supercomputing Centre of the Bavarian Academy of Sciences and Humanities, Garching b.~M\"unchen, Germany. \\ Email: \{fabio.baruffa, luigi.iapichino, nicolay.hammer\}@lrz.de, vasileios.karakasis@cscs.ch}} 


\thispagestyle{plain}
\pagestyle{plain}
\maketitle

\begin{abstract}
We describe a strategy for code modernisation of Gadget, a widely used community code for computational astrophysics. The focus of this work is on node-level performance optimisation, targeting current multi/many-core Intel\textsuperscript{\textregistered}\ architectures. We identify and isolate a sample code kernel,
which is representative of a typical Smoothed Particle Hydrodynamics (SPH) algorithm. The code modifications include threading parallelism optimisation, change of the data layout into Structure of Arrays (SoA), auto-vectorisation and algorithmic improvements in the particle sorting. 
We obtain shorter execution time and improved threading scalability both on \Intel Xeon\textsuperscript{\textregistered}\ ($2.6 \times$ on Ivy Bridge) and Xeon Phi\texttrademark\ ($13.7 \times$ on Knights Corner) systems. 
First few tests of the optimised code result in $19.1 \times$ faster execution on second generation \XeonPhi (Knights Landing), thus demonstrating the portability of the devised optimisation solutions to upcoming architectures. 
\end{abstract}

\section{Introduction}
Among the different branches of physics, astrophysics has a long-standing tradition in the use of numerical simulations.
In particular, in the evolution of the cosmic large-scale structure, one has to deal with a problem which is
inherently multi-dimensional, non-linear, and presents no clear spatial symmetry to be exploited. Although analytical models can
provide some general insight, e.g.\ on scaling relations of galaxy clusters \cite{k86}, this field of study has early moved to an
extensive use of computational models \cite{
s12}. Like in many research areas with a similar evolution, a number of codes widely
used in the community have a development history dating back to the last ten to twenty years. As a consequence, the modernisation
of such codes for their effective use on modern computer architectures has become mandatory. Especially for applications with a wide community, it would be preferable from the user's perspective to optimise an existing code, rather than to develop new numerical tools from scratch.

The computational astrophysics code Gadget \cite{springel2005cosmological} is a cosmological,
fully hybrid MPI + OpenMP parallelised TreePM-MHD-SPH code. In this scheme, both gas and dark matter are discretised by means of particles.
In the TreePM approach, the gravity is split into a long-range part, which is computed by sorting the
particles onto a mesh (PM) and then solving the Poisson equation via FFT methods, and a short-range part,
where a direct sum of the forces between particles is performed. For the short-range part,
forces at intermediate distances can be approximated by grouping particles and taking the moments of these
groups (tree method). 
Additional physics modules use different sub-grid models, to properly treat processes
which are far below the resolution limit in galaxy simulations.

A basic version of the code has been publicly released as Gadget-2 in 2005\footnote{Website: \texttt{http://wwwmpa.mpa-garching.mpg.de/gadget/ } }.
Since then, many groups have started independent development lines (see for example the comparison project described in \cite{syp16}),
some of them leading also to algorithmic improvements (e.~g.~\cite{bdh13}). In this work, we describe a strategy of code modernisation at node level,
performed on the code version dubbed as P-Gadget3 \cite{springel2005cosmological,bma16}.
Our code optimisation work has as target multi-core and many-core Intel\textsuperscript{\textregistered}\ architectures.
We will show that such modern systems expose new performance bottlenecks even on an optimised code which is successfully
used on large-scale HPC systems. In particular, node-level performance plays now a crucial role, because both threading parallelism and
vectorisation have an ever increasing importance for keeping pace with Moore's law. 

We want to stress a basic principle leading our development:
we perform code modifications which are `as non-invasive as possible'. Our intention is to ensure the portability of the code across different 
computer architectures, and the readability for the developers. Furthermore the group of general users, which often have basic programming skills and
are more interested on the physics implementation, have to be able to modify the code without coping with performance questions. 
For all these reasons, our attempts will focus on minimally invasive code modifications, relying on compiler auto-vectorisation and refraining from intrinsics instructions.

In order to have a simpler testbed for our development, an OpenMP-only code kernel has been isolated from P-Gadget3.
The selected kernel is part of the Subfind algorithm (\cite{swt01,dbm09}; Section \ref{subfind}), originally developed as a standalone tool and later embedded into Gadget
for inline data analysis. Besides being representative of analogous code parts in P-Gadget3, this kernel is also relevant
for the vast category of halo finding algorithms in computational astrophysics \cite{bkp15}.

In this paper we focus on the node-level performance optimisation of the Subfind algorithm,
and describe the steps to analyse and improve it. The proposed solutions for removing the performance bottlenecks can be considered as a proof of concept for a future modernisation of the whole code. 
Some otherwise important topics like MPI performance or parallel
I/O go beyond the scope of the present work and will be addressed elsewhere.

The structure of this work is the following: in Section \ref{numerical}, the kernel and its data serialisation are described, and in Section \ref{environment} the hardware and software environment of the tests is presented. The different optimisations applied to the kernel and the related performance improvements are reported in Section \ref{results}. In Section \ref{knl}, it is shown how our strategy of code modernisation improves
the performance on the \Intel Xeon Phi\texttrademark\ of second generation, code-named Knights Landing (thereafter KNL). The conclusions are finally drawn in Section \ref{conclusion}.

\section{Numerical algorithms}
\label{numerical}

\subsection{Particle data and particle interaction scheme in Gadget}
\label{general}

A detailed description of the algorithms and workflow of the Gadget code would go beyond the scope of this paper; we refer the reader to \cite{springel2005cosmological} for a thorough overview of Gadget-2 (whose general structure is largely shared by P-Gadget3). Here we recall only the information which is essential for the current work.

In Gadget, the particles belonging to the same MPI process (so-called local particles) are organised using a Barnes-Hut octal tree
represented as a flat array in memory. The actual particles are not stored in the tree, but just their position in a global particle array per node. This array stores all the local
particles contiguously in memory, with each particle being represented by a large
C \lstinline!struct!. We will get back to the particle data layout and its potential optimisation in Section \ref{particle}.

Both gravitational force computation and search of particle neighbours are performed with the tree algorithm by `traversing' it, in the sense of the hierarchical multipole expansion, implying decisions on opening or not opening tree nodes (\cite{springel2005cosmological} and references therein).
In the following, we refer mainly to the use of tree traversal in the neighbour search: in this case, for any given particle, its neighbours are searched in the tree, and subsequently for each neighbour the particle interactions are calculated.
This particle interaction scheme, consisting of neighbour search and particle interactions, is present in all major code components and comprises most of the total execution time.

\subsection{The Subfind algorithm}
\label{subfind}

Subfind \cite{swt01,dbm09} is a tool for identifying gravitationally bound particle groups (physically representing dark matter haloes) in cosmological simulations, based on the friends-of-friends algorithm \cite{kgk99}. Additionally, Subfind has the capability of detecting gravitationally bound substructures within parent groups. In the detection of substructure, a crucial role is played by the computation of density associated to particles, because the identification of substructures is based on the overdensity with respect to the background value. This computation of density $\rho_i$ at the spatial location $\mathbf{r}_i$ of the particle $i$ is done according to the standard SPH procedure, namely by kernel interpolation over the particle nearest $j$ neighbours:
\begin{equation}
  \rho_i = \rho({\mathbf{r}_i})= \sum_j m_j W(|\mathbf{r}_i -\mathbf{r}_j|, h_j)\,\, .
\label{sph}
\end{equation}
In (\ref{sph}), $m_j$ is the mass associated with the particle $j$, $W$ is the kernel function, and $h_j$ is the kernel smoothing length for particle $j$. In our runs, the Wedland C6 kernel \cite{da12} will be adopted as kernel function, using 295 neighbours.

Equation (\ref{sph}) is implemented in the code (Listing \ref{loop-implemented}) as a \lstinline!for! loop over the \texttt{n} neighbours of a given particle. For every \texttt{n}, the index \texttt{j} in the particle data structure is retrieved in line 2 (implications for the memory access of this data structure are discussed in Section \ref{particle}), and in line 3 the distance of particle \texttt{n} from the given particle is checked. For a neighbour particle \texttt{n} within a smoothing length from the given particle, two inlined functions are evaluated. They return as output a value for the kernel function $W$ (\texttt{w} in line 6), which is then used for the density sum. We stress here that the code described in Listing \ref{loop-implemented} finds application not only in the Subfind algorithm but in several instances of Gadget, therefore its performance optimisation is of general interest for a wide variety of SPH codes. Moreover, as pointed out in \cite{rtb16}, this algorithm has access patterns which are pretty similar to those arising from the use of Verlet lists in Molecular Dynamics, hence this work is relevant even beyond astrophysics.

\begin{lstlisting}[float,caption={Pseudocode for the density computation (Equation \ref{sph}), typical of SPH schemes. Code and variables are described in the text.},label={loop-implemented}]
for (n = 0, n < neighbouring particles){
  j = ngblist[n];
  if (particle n within smoothing length){
    inlined_function1(..., &w);
    inlined_function2(..., &w);
    rho += Particle[j].Mass * w;
  }
  // other particle interaction computations
}
\end{lstlisting}

From the Subfind algorithm, only the kernel \texttt{subfind\_density}, responsible for the density computation described above, has been isolated through the serialisation procedure, motivated and discussed in the next section.
This kernel presents both phases of the particle interactions introduced in Section \ref{general}. In every Gadget component, the workload fraction of the two phases is different. The first phase is nearly the same for all the components, while the second one varies from kernel to kernel, depending on the physics interactions to be calculated.  For \texttt{subfind\_density} the second phase has less workload, but on the other hand it consists mainly of the code in Listing \ref{loop-implemented}, which has a fundamental importance in every SPH implementation. Because of this, we decided to focus on the optimisation of this kernel, even though in this case the workload of the particle interactions in Listing \ref{loop-implemented} is subdominant with respect to the neighbour search (cf.\ Section \ref{particle}).

\subsection{Isolating a representative code kernel}
\label{serialisation}

The key concept behind the isolation of \texttt{subfind\_density} is a simple serialisation library, implemented as a wrapper to the C standard libraries \texttt{fwrite()} and \texttt{fread()}, adding additional functionality, such as checksumming. The isolated \texttt{subfind\_density} can be then invoked independently of the main application, but with the exact same input and environment as it had originally. The first step to achieve this is to identify which data is actually accessed by the routines of interest and dump them on the disk, to be later used `as-is' by the isolated kernel.
For the Subfind kernel one has to dump the complete particle array, the particle tree data structures and the global configuration variables (held in a global data structure), plus a few additional variables. This sums up to $551$\ MB for the physics workload under examination (a simulation evolving $81^3$ dark matter particles and the same number of gas particles).

As a design choice, the isolated kernel has been implemented within a simple front-end driver, completely integrated with the build framework of Gadget. The driver deserialises the input data from the disk, feeds the kernel and finally dumps and verifies the result. By this it simulates the original algorithm behaviour, as far as data structure, variable and function definitions are concerned. The resulting quick prototyping and testing of different optimisation solutions is of great help and importance in the code modernisation process. Moreover we notice that, for simplicity and for focusing on the node-level performance, every reference in the original source code to the MPI parallelism has been removed from the kernel, reducing it to pure OpenMP shared memory parallelism.

\section{Systems and software environment}
\label{environment}

The results presented here are based on tests performed on the SuperMIC cluster at
LRZ\footnote{\texttt{https://www.lrz.de/services/compute/supermuc/\-supermic/}}. The system is equipped with 32 nodes,
each containing two 8-core \Intel Xeon\textsuperscript{\textregistered}\ E5-2650v2 processors (code-named Ivy-Bridge, henceforth IVB) @ 2.6\ GHz. 
To each host node, two \Intel \XeonPhi Knights Corner (KNC) coprocessors 5110P with 60 cores @ 1.1\ GHz are attached.
For all KNC tests presented here, the kernel is executed in native mode.

We use the Intel\textsuperscript{\textregistered}\ C++ Compiler 16.0. Compilation flags, if not otherwise specified, are:
\begin{itemize}
\item \texttt{-xhost} for generating vectorised code on IVB;
\item \texttt{-mmic} for generating vectorised code on KNC.
\item \texttt{-no-vec -no-simd} for scalar code on IVB;
\item \texttt{-mmic -no-vec -no-simd} for scalar code on KNC;
\item In all the cases, we use the compiler flags \texttt{-O2 -qopenmp}.
\end{itemize}

For the multithreaded version of the code (OpenMP), the thread execution has been bound to physical processing units (thread pinning) using the thread affinity interface of the Intel\textsuperscript{\textregistered}\ OpenMP$^*$ Runtime Library. 
On IVB the assignment of consecutive threads is done on physical cores which are close in topology to each other, using multiple assignments (simultaneous multithreading, SMT) only if all physical cores are filled. On KNC an even distribution throughout the cores has been guaranteed\footnote{In terms of the environment variable \texttt{KMP\_AFFINITY}, these assignments (tuned by optimal performance) are equivalent to \texttt{granularity=fine, com\-pact,1,0} on IVB and to \texttt{granularity=fine,scatter} on KNC.}.
SMT is used both on IVB
(two threads per core, resulting in tests with 32 threads) and on KNC,
where up to four threads per core (a maximum of 240 threads in total) are available. Especially on this system, the efficient utilisation of multiple threads per core is crucial for getting performance \cite{colfax}.

The application analysis is performed using tools from
the \Intel Parallel Studio XE 2017. In particular, the multithreading issues have been identified using VTune\texttrademark \ Amplifier XE, and the vectorisation prototyping has been investigated with Intel\textsuperscript{\textregistered}\ Advisor. Furthermore, when generating vectorised code we make extensive use of the vectorisation reports, 
 produced with the compiler flag \texttt{-qopt-report=5}. The reports provide a good insight on the optimisations performed automatically by the compiler. In Section \ref{particle}, the cache performance has been studied with LIKWID \cite{thw11}.

\section{Performance results}
\label{results}

\subsection{Threading parallelism}
\label{lockfix}

Before focusing on the particle interaction phase,  the OpenMP implementation of the \texttt{subfind\_density} kernel (similar to that of other components of P-Gadget3) will be analysed and improved in this section.
The version of the kernel we start our optimisation from, thereafter dubbed as \textit{original}, has been profiled with 8 threads using \VTune Amplifier XE on an IVB node. The profiling using the analysis type \textit{Basic hotspots} shows a severe OpenMP parallelisation overhead, with more than 40\% of the execution time spent in threads spinning on locks. 


\begin{lstlisting}[float,basicstyle=\ttfamily\small,caption={Parallel particle interaction scheme in Gadget (pseudocode in C++11 notation).},label={alg-subfind-orig}]
more_particles = partlist.length;
while (more_particles) {
  int i = 0;
  while (!error && i < partlist.length) {
   #pragma omp parallel
   {
     #pragma omp critical
     {
       p = partlist[i++];
     }
     if (!must_compute(p))
       continue;
     ngblist = find_neighbours(p);
     sort(ngblist);
     for (auto n : select(ngblist, K))
       compute_interactions(p, n);
   }
  }
  more_particles = mark_for_recomputation(partlist);
}
\end{lstlisting}

Taking a closer look at the original algorithm (schematically drawn in Listing~\ref{alg-subfind-orig}), it is easy to spot the potential bottleneck at the lock set when accessing the particle list. More specifically, within the list of particles \texttt{partlist}, each thread gets
 the next particle to process within the \texttt{critical} code section (line 7), and checks only then whether it is actually needed to be processed (line 11).
If this is the case, then it proceeds with the neighbour finding and particle sorting (lines 13 and 14; sorting will be discussed in detail in Section \ref{qselect}) and the particle interaction phase (line 16, basically the code in Listing \ref{loop-implemented}), otherwise tries to get the next particle in the list.
The reason behind this is that, at the end of each iteration on particles, a set of them is marked for recomputation (line 19), if they have a number of neighbours smaller than 295. For these particles, the smoothing length is adjusted, and the algorithm is repeated until no particles are left. The number of particles to be recomputed in each iteration decreases exponentially, almost halving in every iteration. For our workload, a total of 18 iterations of the outermost loop (line 2) is required. The problem with the locking scheme of this algorithm is that, at later iterations, the particle list is locked and unlocked constantly since most of the particles need not be recomputed, creating a tremendous parallelisation overhead. 

Having identified the problem, one can devise more than a single solution. In a first optimisation attempt,
a `todo' particle list, which contains only particles that need to be recomputed in each iteration, has been introduced.
Instead of iterating the whole particle list every time, the iteration is performed only over this `todo' list.
This intermediate solution 
fixes the problem of unnecessary locking and
unlocking the particle list, since it is guaranteed that the acquired particle needs to be recomputed.
This new list has to be updated at the end of each iteration, but this incurs negligible overhead,  since it can be easily integrated in the process of marking the particles for recomputation. This improved strategy is quite efficient on IVB, but still shows limitations for very large thread counts on KNC. The key efficiency problem still stems from the common particle queue, from which all threads try to fetch the next particle to compute.

A tempting approach for better fixing this problem is to split the particle queue into equal parts,
and assign each part to a separate thread for processing. This approach, however, is
problematic since the work associated with each particle is not known beforehand. Indeed, trying to implement such a
static partitioning scheme leads to severe load imbalance among the threads and worse performance than the version with lock contention fix.
In order to deal with the unavoidable intrinsic variations in the amount of work associated with each particle (caused for instance by accessing neighbours from memory), the individual work items must be scheduled dynamically among the threads. The most straightforward way to implement this takes advantage of the OpenMP's work sharing capabilities, by means of a simple \lstinline!omp for! loop using dynamic scheduling (Listing \ref{alg-subfind-lockless}). We notice here at line 1 the particle list \texttt{todo\_partlist}, consisting of particles flagged for being recomputed, as described above.
Additionally, to suppress lock contention completely, the critical section inside the loop must be avoided. To achieve this,
we have restructured the loop by
moving all premature exit condition checking (compare Listing \ref{alg-subfind-orig}, line 4, with Listing \ref{alg-subfind-lockless}, lines 6--7) outside the loop (i.e., at line 13 of Listing \ref{alg-subfind-lockless}).

\begin{lstlisting}[float,caption={Like Listing \ref{alg-subfind-orig}, but with the work sharing implemented through a \lstinline!omp for! loop with dynamic scheduling. },label={alg-subfind-lockless}]
todo_partlist = partlist.length;
while (todo_partlist) { 
  error = 0;
  #pragma omp parallel for schedule(dynamic)
  for (auto p : todo_partlist) {
    if (something_is_wrong) 
      error = 1;
    ngblist = find_neighbours(p);
    sort(ngblist);
    for (auto n : select(ngblist, K))
      compute_interactions(p, n);
  }
  // ... check for any errors
  todo_partlist = mark_for_recomputation(partlist);
}
\end{lstlisting}

We will dub the kernel version described by Listing \ref{alg-subfind-lockless} as \textit{lockless} hereafter, because the explicit locks are removed (although they may still be used by the OpenMP implementation). The improvement of performance produced by this new work sharing strategy is shown firstly by the dramatic decrease of OpenMP overhead: 
the fraction of execution time spent on thread spinning in this improved version is $3\%$, to be compared with $44\%$ of the \textit{original} version.


\begin{figure}
\centering
\includegraphics[width=0.95\linewidth]{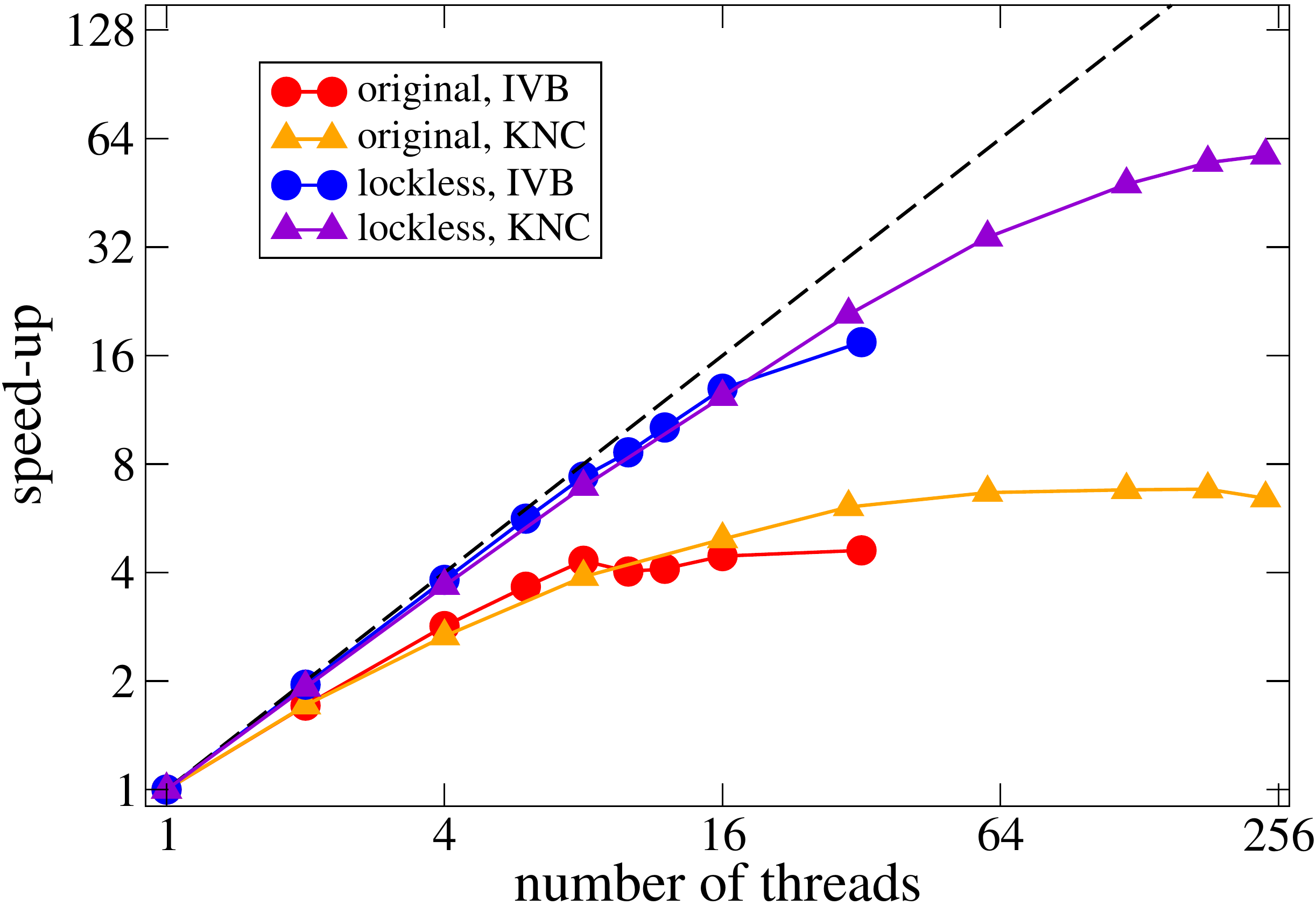}
  \caption{Parallel speed-up of the \texttt{subfind\_density} kernel on IVB (lines with circles) and KNC (lines with triangles), in a comparison of the \textit{original} and \textit{lockless} code versions. The different runs are indicated by lines and symbols as in the legend, while the dashed black line denotes ideal speed-up. All data are relative, and normalised to the one-thread performance of the corresponding architecture.}
  \label{fig.speedup.lockless}
\end{figure}

Another useful information about the performance improvement can be obtained by a thread scaling analysis of the kernel (Figure \ref{fig.speedup.lockless}). In the \textit{original} kernel version,
the lock contention problem is also the source of a slowdown observed on the scaling plot with increasing thread number.
In particular, moving beyond a single socket on IVB has a large negative impact on performance, since locking cost becomes
quite expensive when synchronising across sockets in NUMA systems. Conversely, the code optimisation in the \textit{lockless}
version increased the scalability of the kernel, especially on the Xeon Phi.

More specifically, the speed-up has improved by $1.8 \times$ on a IVB socket (8 threads), reaching  $92\%$
parallel efficiency, and a by significant $5.2 \times$ on KNC with 60 threads ($57\%$ parallel efficiency after the optimisation).
Important to note, the kernel performance in the \textit{lockless} version can efficiently scale
across the sockets. Furthermore, it can benefit from the use of SMT on IVB, which was not the case for the \textit{original} version.
In addition on KNC, with the \textit{lockless} version the performance is improved also when moving beyond one thread per core.

Since the single thread performance has not changed, these results automatically translate into a time
to solution speed-up. Concerning the performance comparison between \Xeon and Xeon Phi, for the \textit{original} kernel version KNC was $6.25 \times$ slower than IVB, while for \textit{lockless} kernel it is only $2.1 \times$ (comparing 60 threads on KNC and 8 threads on IVB). Optimising the work sharing among the OpenMP threads has thus allowed KNC performance to nearly meet half of the performance of a single IVB socket.

\subsection{Enabling vectorisation: redesign of the particle data layout}
\label{particle}

After having improved the work sharing strategy of  \texttt{sub\-find\-\_density}, the next logical step is to investigate its potential for vectorisation. However a performance bottleneck on the data structure has to be removed first. This section is dedicated to this topic.

In order to obtain timing information and vector optimisation hints at loop level, the \textit{lockless} version is
compiled on IVB and profiled using \Intel Advisor.
We observe in the \textit{Loop Metrics} that $99\%$ of the total runtime is spent in scalar loops. Among them, the most expensive part in terms of runtime ($83\%)$ is given
by the tree-walk computation needed for the neighbour search. Its current implementation involves a \texttt{while} loop,
which is not a suitable candidate for being automatically vectorised by the compiler. 
A modification of this part of the algorithm would be a major code restructuring, which is out of the scope of the current work.
However, an essential part of the neighbour finding phase is the particle sorting, which is subject of the algorithmic improvement in Section \ref{qselect}. 

The second most expensive loop ($13\%$ of the total runtime) is instead more suitable for our vectorisation analysis. This corresponds to the computation described in Listing \ref{loop-implemented}, whose general importance has been introduced in Section \ref{subfind}.
In the \textit{lockless} kernel version, this loop is not vectorised due to the data layout of the code,
which prevents spatial data locality. For this reason, before dedicating any effort to vectorisation, an essential prerequisite is a redesign of the data structure.

As mentioned in Section \ref{general}, in P-Gadget3 the particle data are organised as a C \lstinline!struct!, 
and the collection of all of them is implemented as array of structures (AoS for brevity).
This kind of data organisation is typical for particle codes 
because it naturally allows the addition of further physical quantities into the algorithmic scheme with minimal coding overhead.
In P-Gadget3, and consequently in \texttt{subfind\_density}, the particle data structure can be quite large,
especially in problems involving a large number of different physical variables. In our test case, the particle data structure object has indeed 224 bytes per particle.

It is well known that accessing a particle data structure organised as AoS hinder efficient cache utilisation.      
In the case of \texttt{subfind\_density}, 
a \VTune Amplifier XE \textit{General exploration} analysis on IVB shows that the kernel is back-end bound, a situation typically hinting to data cache misses.
The very large particle \lstinline!struct! incurs memory accesses with large strides, even if the neighbours are accessed completely sequentially in the particle array. 

Moreover the utilisation of Single Instruction Multiple Data (SIMD) registers is suboptimal with this data layout. According to the general auto-vectorisation guide \cite{autovec}, the \Intel C++ Compiler 
can automatically generate a vectorised code version for loops computing on AoS. Presumably,
the performance of such code is very low, due to the non-unit strided memory access. However in our specific case the loop was not even vectorised by the compiler, because of the unfavourable data layout.

The reasons given above motivate the change of the data structure into a structure of arrays (henceforth SoA),
in order to promote better data locality and auto-vectorisation, and thus to explore the vector capabilities of the tested hardware. 
For this change of data layout, a minimally invasive approach has been chosen: 
we assume that the code framework surrounding the kernel is still making use of the AoS data structure implementation (e.g. for the MPI
communication), and that the new SoA data structure is only exposed in loops over particle neighbours. This choice thus restricts the code modifications that are required, and makes the consistency checks simpler, at the cost of additional routines for copying data from AoS to SoA and vice-versa.

Among the many variables defined in the former particle data structure \texttt{Particle} (cf.\ Listing \ref{loop-implemented}, line 6), 
only those used in \texttt{subfind\_density} are converted into the new
SoA particle data structure \texttt{ParticleSoA}. This results in a more compact particle data structure (60 bytes per particle).
We implemented software gather and scatter routines, in order
to keep the new improvement consistent with the kernel infrastructure. The gather
function is called before the computation to copy the necessary data from \texttt{Particle} into \texttt{ParticleSoA}. 
After the computation, using the SoA layout, the data is copied back (scatter) to the AoS \texttt{Particle}. 
The gather and scatter routines include threading parallelism. 
The overhead of these routines sum up to at most $1.8 \%$ of the total runtime, both on IVB and KNC.
We dub the resulting new kernel version as \textit{SoA}. The performance improvements introduced in this version are clearly visible from the better memory data access. 
In the initial analysis using LIKWID we observe a large fraction of cache misses ($33\%$), which are lowered to $19\%$ when the SoA particle data structure is used.

\begin{figure}
  \centering
  \includegraphics[width=0.95\linewidth]{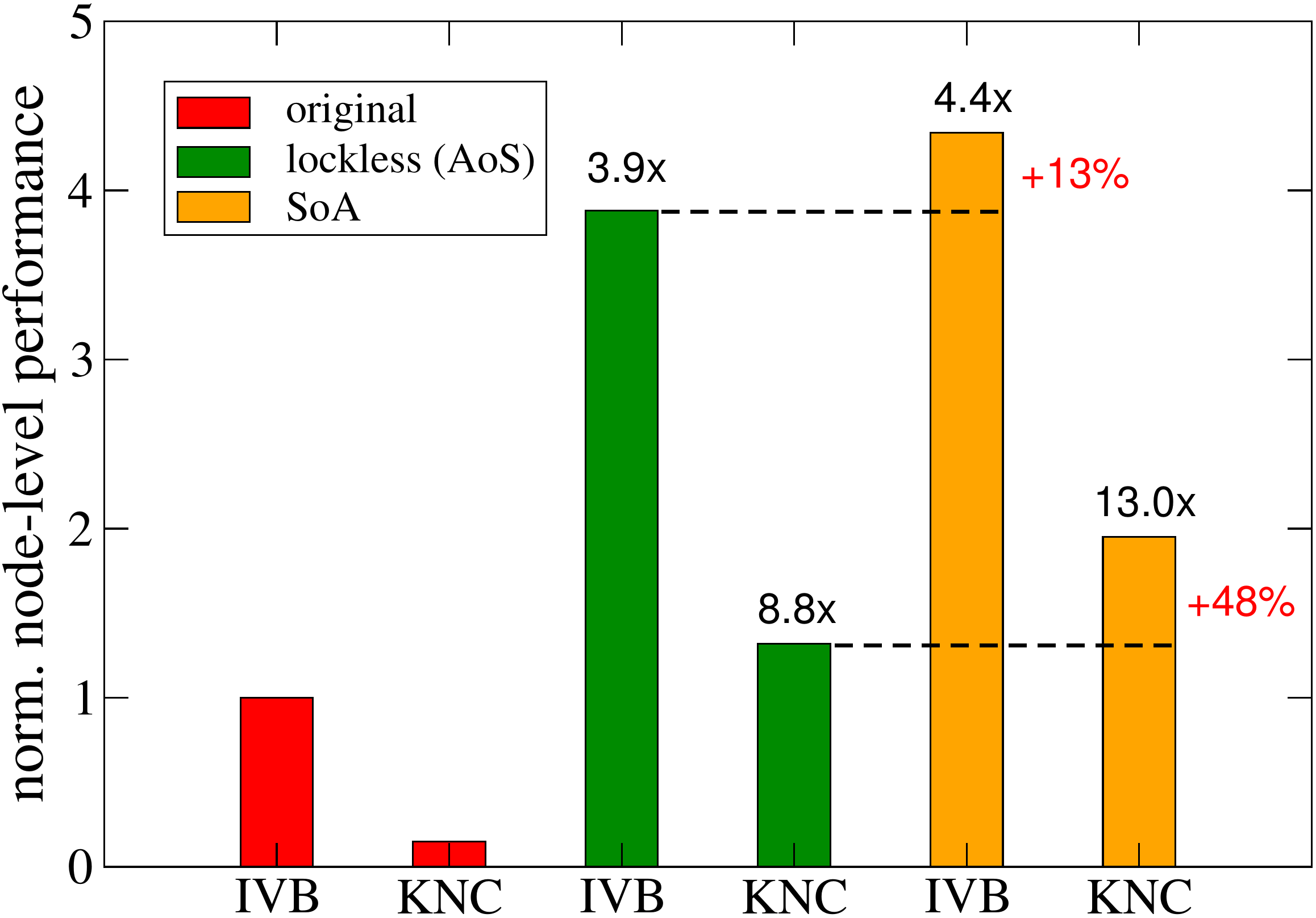}
  \caption{Node-level performance, defined as the inverse of the execution time, on IVB and KNC at three different stages
  of our optimisation process, as indicated in the legend. The performance is normalised to the best original IVB run (bar on the left-hand side);
  higher bars correspond to higher performance.}
  \label{histo}
\end{figure}

An overview of the node-level performance improvements, including the benefit associated with the introduction of the SoA, is shown in Figure \ref{histo}. 
Here the inverse of the execution time is shown and the best result on the IVB node (or on the KNC) is reported,
including SMT whenever applicable. Changing the data structure to SoA gives an additional $13\%$ improvement in performance on IVB, while
a significant impact ($48\%$) is apparent on KNC. Compared with the gain between the \textit{original} and the \textit{lockless} version, 
the improvement introduced by the \textit{SoA} version seems less remarkable (leaving aside the benefit for memory access).
However, the main goal for introducing the SoA data structure, namely enabling auto-vectorisation, has been achieved. 
The results in Figure \ref{histo} still refer to the kernel compiled for generating scalar code (i.e.~options \texttt{-no-vec -no-simd}). 
Results for vectorised code are discussed in Section \ref{vectorisation}.

\subsection{Optimising vectorisation}
\label{vectorisation}

With the change in data layout described in the Section \ref{particle}, the compiler is finally able to automatically vectorise our target loop, previously identified as the one shown in Listing \ref{loop-implemented}.
Now we want to evaluate the speed-up introduced by vectorising this loop on different systems, and possible further improvements.

From the SIMD perspective, probably the most problematic point in Listing \ref{loop-implemented} 
is the relatively large part embraced in the \texttt{if} statement starting at line 3. 
If the compiler implements the \texttt{if} condition through masking, all subsequent statements inside the condition (and function calls) need to be masked, too. To reduce the masking overhead, one should put the condition as low as possible in the call tree. Therefore we have moved the \texttt{if} statement inside \texttt{inlined\_function1}. When the \texttt{if} condition is not fulfilled, \texttt{w = 0.0} is returned in line 6 of Listing \ref{loop-implemented}, making this implementation equivalent to the initial one. 
This new kernel version is dubbed as \textit{vectorised}.

We measure the runtime spent in the target loop to quantify the vectorisation speed-up, 
defined as the ratio between the time spent in the scalar loop and the vectorised versions:
\begin{equation}
S_\textrm{v} = \frac{t_{\textrm{loop},\ \texttt{-novec -nosimd}}}{t_{\textrm{loop},\ \texttt{-xhost}}} \,\, ; \, \, \, \epsilon = \frac{S_\textrm{v} }{VL}  \, \, .
\end{equation}
The corresponding (architecture-dependent) vector efficiency $\epsilon$ is defined as the ratio of $S_\textrm{v}$ and $VL$, which is the 
loop vector length. For the  loop under investigation, $VL =4$ on IVB and 8 on KNC, since double precision FP values are used in the computation. 

In the \textit{SoA} version the loop was vectorised, as confirmed by the compiler report. However no performance benefits with respect to the 
scalar code are measured, neither on IVB nor on KNC, resulting in $S_\textrm{v} \simeq 1\times$. An improvement is observed between the 
kernel versions \textit{SoA} and \textit{vectorised}, which supports our assumption of a masking bottleneck. More specifically, 
for IVB $S_\textrm{v} = 2.2\times$ ($\epsilon = 55\%$), while for KNC $S_\textrm{v} = 3.4\times$ ($\epsilon = 42\%$). 
Although very promising, the reported speed-ups are still well below the ideal vector speed-up. 
This is due to several reasons. First, the use of auto-vectorisation is generally less efficient than programming with intrinsics.
Second, the way the particle data structure is accessed (line 2 in Listing \ref{loop-implemented}) leads to strided memory access, 
as also been confirmed by the \textit{Memory Access Pattern} analysis of \Intel Advisor.
Furthermore additional optimisation like memory-aligned data access is missing (as shown by the compiler report) 
and definitely deserves attention in future work.

\subsection{Further algorithmic improvements}
\label{qselect}

Examining again the execution time profile of \texttt{subfind\_density} with the previous optimisations reveals that a significant amount of time ($\sim 50\%$)
is spent sorting the neighbour list.
The original algorithm (cf.\ Listing \ref{alg-subfind-lockless}, line 9) first sorts the resulting neighbour list according to their distance from the 
current particle and then selects the first $K$ neighbours ($K = 295$ in our configuration). 
However, in this specific kernel the sorting information is not used beyond that point, meaning that the only purpose of sorting is to select the $K$ 
nearest neighbours from the neighbour list.
This selection can be achieved in linear scaling time with the number $N$ of particles, instead of the $O(N\log N)$ sorting time, by only partially sorting the neighbours.

An array of $A$ elements, on which an ordering relation is defined, can be partially sorted in linear time, so that $A_i \leq A_K$ for each $i \leq K$.
For our purpose, we have used the QUICKSELECT algorithm \cite{quickselect}, which is based on QUICKSORT.
Although its worst-case performance is of $O(N^2)$ as in case of QUICKSORT, in the average case it is of $O(N)$. Therefore, it is a good candidate for substituting the complete sorting in our case.
Our initial experiments comparing the two algorithms (our straightforward QUICKSELECT
implementation vs.\ C's QUICKSORT) showed that the difference in average complexity pays off for very large arrays where the
performance difference is at least one order of magnitude.
However, the neighbour list that needs to be sorted in the Subfind kernel is relatively small (500--1000 particles), but this operation is executed nearly a million times in our test.
The small neighbour list tightens the margin of the performance improvement we get by replacing QUICKSORT in the Subfind kernel.
However, the replacement still brings to a significant improvement of 41\%
on IVB (16 threads) and 27\%
on KNC (120 threads) compared to the \textit{vectorised} kernel version described in the previous section. The kernel version discussed in this section represents the final step of the optimisation process presented in this work, and will be henceforth indicated as \textit{optimised}.

\section{Performance on \Intel Knights Landing}
\label{knl}

The optimisations described in Section \ref{results} have the merit of being
inspired by a general approach (\textit{code modernisation}, \cite{code-mod}), which can be used for
porting and improving the performance on modern computer architectures.
In this section we verify that the code improvements described in this work lead to a reproducible performance gain 
also on newer hardware, which was not the original target of our project. More specifically, we show here results of tests performed on \Intel \Xeon processor
E5-2697v3 (code-named Haswell, HSW), with 28 cores @ $2.3$\ GHz, and on the \Intel \XeonPhi 7210 with 64 cores @ $1.3$\ GHz (code-named Knights Landing, KNL).

The software environment is the same as described in Section \ref{environment}. For HSW, the same compilation flags of IVB were used. In the case of KNL, a full exploration of memory and cluster modes available on this system \cite{va16,asai16b} would go beyond the scope of this paper, and will be subject of forthcoming work. Our settings are \textit{Flat} for memory mode and \textit{All--to--All} for cluster mode. The compilation flags on KNL are: \texttt{-xMIC-AVX512 -O2 -qopenmp}. For simplicity of use, we decided to allocate on the High-Bandwidth Memory as preferred option, and falling back to DDR memory if needed\footnote{This is achieved at runtime by \texttt{numactl -preferred 1}.}. The thread pinning has been set similarly to KNC (\texttt{KMP\_AFFINITY=scatter}).

\begin{figure}
\centering
\includegraphics[width=0.95\linewidth]{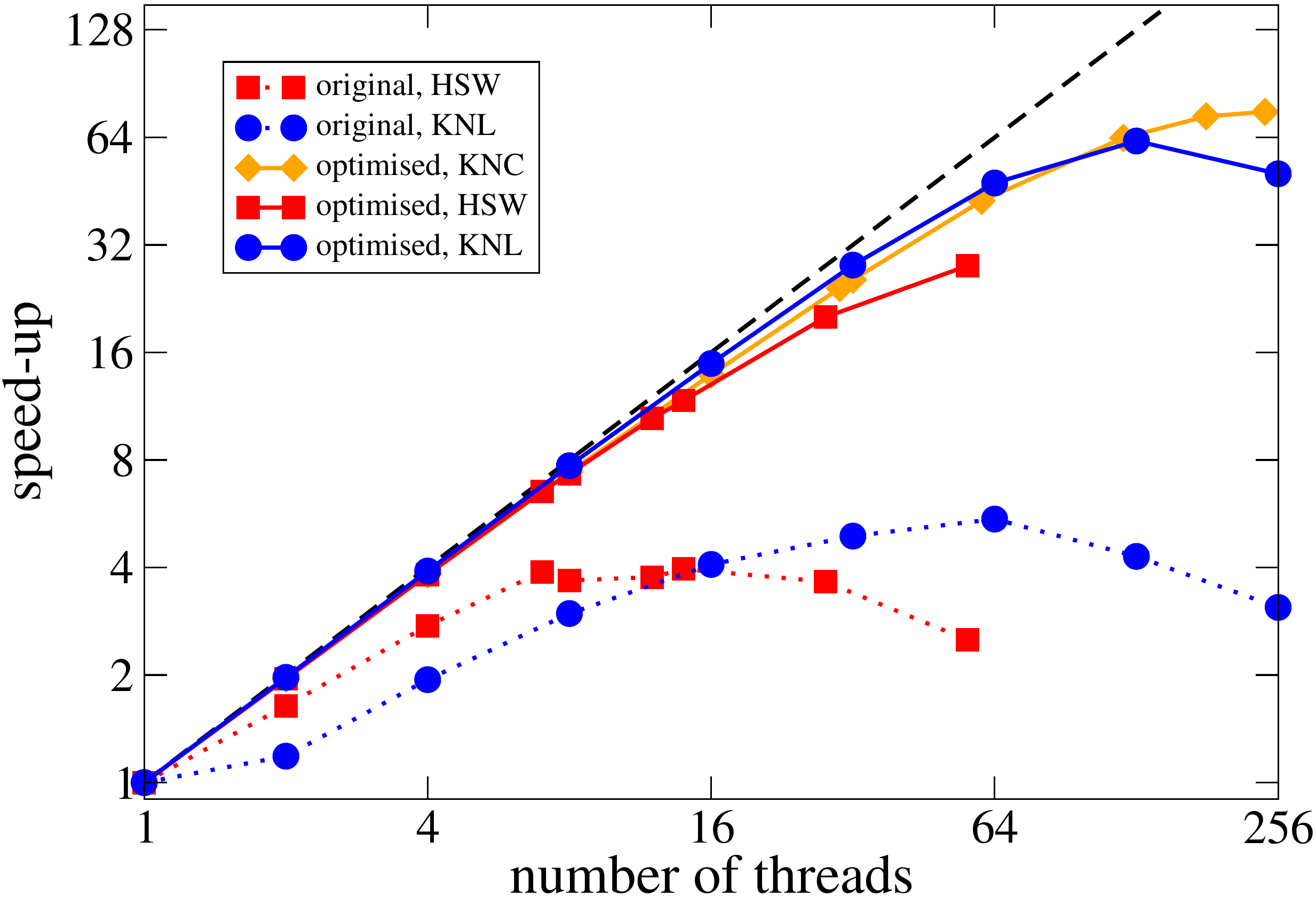}
  \caption{Parallel speed-up of \texttt{subfind\_density} on different architectures (see legend). The dotted lines
  correspond to the \textit{original} version of the kernel, while the solid lines to the \textit{optimised} version. All data are relative, and normalised to the one-thread performance of the corresponding architecture.}
  \label{perf_final}
\end{figure}

In Figure \ref{perf_final}  the scaling with the number of threads of the \textit{optimised} version of the kernel is compared with the  \textit{original} version, for HSW and KNL. As a term of comparison, the line for the  \textit{optimised} KNC speed-up is added.

The \textit{original} kernel shows poor scaling also on the newest architectures, similar to what initially seen in Figure \ref{fig.speedup.lockless}. The \textit{optimised} version, on the other hand,
reproduces the performance improvement observed on IVB and KNC.
Between the two versions we measure a speed-up of $2.9 \times$ on HSW (14 threads), and a significant $8.7 \times$ improvement on KNL (64 threads). On more than two threads per core the scaling for KNL is lower than the one for KNC, indicating a different architectural need for SMT between the two generations of Xeon Phi. Final results on  time to solution will be presented in the next section, but we mention here for completing this discussion that the single-thread execution time on KNL is $3.3 \times$ faster than on KNC for the \textit{optimised} kernel.

Finally, the auto-vectorisation of the target loop (Section \ref{vectorisation}) has a remarkable performance on KNL, where we measure $S_\textrm{v} = 6.62 \times$ ($\epsilon = 83\%$, having $VL = 8$). The reason for the improvement over IVB and KNC has not been explored yet. We speculate that it might be caused by the use of the Intel\textsuperscript{\textregistered} AVX512 instruction set,
 featuring more efficient gathers and mask manipulation instructions.

\section{Conclusions and future work}
\label{conclusion}

Many community codes have been originally designed as a serial application, and grown by merely adding different layers of parallelism. 
Code modernisation is important for these codes, in order to be prepared for next generation processors, and to immediately  profit from the current hardware.
In this work we presented node-level performance optimisation of algorithms used in P-Gadget3,
having as target \Intel \Xeon and \XeonPhi architectures. The described strategies and the resulting code modifications, such as prevention of the lock contention, SoA data layout and vectorisation improvement are general, and applicable
to a broad range of similar codes. The performance improvements are significant both in terms of threading scalability
(Figures \ref{fig.speedup.lockless} and \ref{perf_final}) and execution time. 

\begin{figure}
\centering
\includegraphics[width=0.95\linewidth]{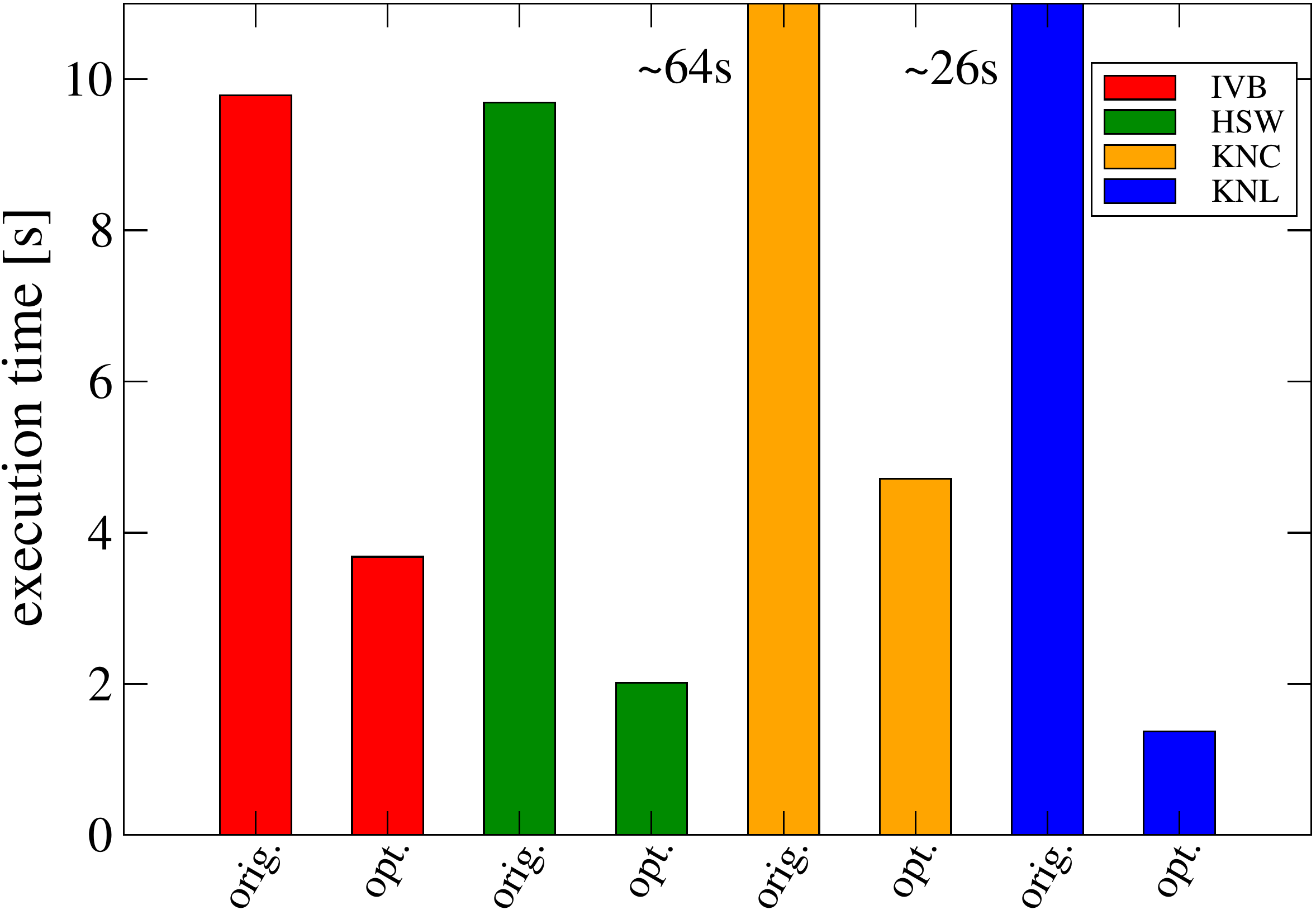}
  \caption{Execution time of  \texttt{subfind\_density} on different architectures, indicated in the legend, for the kernel versions \textit{original} and \textit{optimised}. The data refer to tests performed on 8 threads (one socket) for IVB, 14 threads (one socket) for HSW, 240 threads (i.e.\ four threads per core) for KNC and 128 threads (two threads per core) for KNL.}
  \label{perf_bar}
\end{figure}

Concerning these final diagnostics, a summary of obtained results on all tested architectures is shown in Figure \ref{perf_bar}. 
We can see an improvement in runtime by factors of $2.6$ and $4.8$ on IVB and HSW, respectively. It has been verified that performance on 
the latter system benefits also from the compiler generation of Fused Multiply-Add (FMA) instructions. 

The unsatisfactory result of the \textit{original} kernel on KNC highlights a known feature of this system, 
namely its sensitivity to parallelism bottlenecks. On the other hand, KNC benefits considerably (and more than Xeon) 
from the code improvements, with a total speed-up of $13.7 \times$. 
The performance gap between KNC and IVB then decreases from $6.6 \times$ to $1.3 \times$.
The results for KNL show that this system is more tolerant concerning performance issues than KNC.
Indeed, on KNL the \textit{optimised} kernel is $19.1 \times$ faster than \textit{original}, and $1.5 \times$ faster than the result obtained on the HSW architecture. 
We stress once more that KNL, despite of its good performance, was not the original target of this project. Early work on Intel Xeon E5-2699v4 (Broadwell) further confirms these findings (total speed-up: $4.7 \times$) and thus the portability of the presented approach.

The relevance of the presented work goes beyond \texttt{subfind\_density}, and impacts the modernisation of the Gadget code and of similar community applications. Our path for minimally invasive, portable and readable optimisation is a clean and straightforward strategy, for the wide Gadget community to be adopted in their own development lines.
Future work will certainly include a plan to backport our optimisation solutions to the most important parts of P-Gadget3. 
This task is demanding, because it is going to touch the current MPI implementation of the code and its underlying data structure, defined as AoS. 
Moreover, after the first tests shown here, more work will be devoted to KNL-specific features (use of MCDRAM, Intel AVX512 instruction set, clustering modes) 
and their potential for optimisation. On the long term, one must not exclude further
modifications, like refactoring the tree traversal algorithm, in a fashion which is more favourable for execution on many-core systems with large vector registers.

\subsubsection*{Acknowledgement} The research of F.~B., L.~I. and V.~K.~
has been partly supported by the Intel Parallel Computing Center (Intel PCC) \textit{ExScaMIC -- Extreme Scaling on MIC/x86} at LRZ and Technical University of Munich (TUM).
Thanks to N.~Tchipev (TUM) for valuable suggestions on the draft, to K.~Dolag, M.~Petkova and A.~Ragagnin for their support in the use of Gadget,
and to H.~Bockhorst and G.~Zitzlsberger (Intel) for consulting and careful reading of the manuscript.
Intel, Xeon, Xeon Phi and VTune are trademarks of Intel Corporation or its subsidiaries in the U.S. and/or other countries.

\bibliography{IEEEabrv,report-index}
\bibliographystyle{IEEEtran}

\end{document}